\documentclass[lettersize,journal]{IEEEtran}
\usepackage[table,dvipsnames]{xcolor}
\usepackage{placeins}
\usepackage{float}
\usepackage{amsmath,amsfonts}
\usepackage{amssymb}
\usepackage{algorithmic}
\usepackage{array}
\usepackage[normalem]{ulem}
\usepackage[caption=false,font=normalsize,labelfont=sf,textfont=sf]{subfig}
\usepackage{textcomp}
\usepackage{stfloats}
\usepackage{url}
\usepackage{verbatim}
\usepackage{cite}
\def\BibTeX{{\rm B\kern-.05em{\sc i\kern-.025em b}\kern-.08em
    T\kern-.1667em\lower.7ex\hbox{E}\kern-.125emX}}
\usepackage{balance}

\usepackage{graphicx}
\graphicspath{{./Figs/}}
\DeclareGraphicsExtensions{.pdf,.jpeg,.png,.eps}

\usepackage{booktabs}
\usepackage{xurl}
\usepackage{multirow}
\usepackage{tabularx}
\usepackage{siunitx} 
\usepackage{caption}
\usepackage{amsfonts}
\usepackage{dcolumn}
\usepackage{xspace}

\usepackage[T1]{fontenc}
\usepackage[utf8]{inputenc} 
\usepackage{listings}

\usepackage{hyperref}
\usepackage{pbox} 

\usepackage{seqsplit}
\usepackage{csquotes}
\usepackage{enumerate}
\usepackage[most]{tcolorbox}
\usepackage[export]{adjustbox}
\usepackage{tikz}
\usepackage{makecell}
\usepackage{threeparttable}
\newcolumntype{L}[1]{>{\raggedright\arraybackslash}m{#1}} 
\newcolumntype{R}[1]{>{\raggedleft\arraybackslash}m{#1}}  
\newcolumntype{Y}{>{\raggedright\arraybackslash}X}        

\newcommand{\citep}[1]{\cite{#1}}

\newcommand{\citet}[1]{\csuse{mapcitet#1}~\cite{#1}}

\definecolor{custom-gray}{cmyk}{0,0,0,0.7,1.00}
\definecolor{rank1}{HTML}{99FF99}
\definecolor{rank2}{HTML}{FFFF99}
\definecolor{rank3}{HTML}{FFCC99}
\definecolor{rank4}{HTML}{FF9999}
\definecolor{rank5}{HTML}{DDDDDD}  
\newtcolorbox{Summary}[2][]{
    enhanced,
    sharp corners,
    colback=white,
    colframe=black,
    boxrule=0.4pt,
    left=4pt,
    right=4pt,
    top=3pt,
    bottom=3pt,
    boxsep=0pt,
    before skip=4pt,
    after skip=4pt,
    fontupper=\normalsize,
    #1,
    before upper={\textbf{#2.}\ }
}

\lstdefinelanguage{json}{
  upquote=true,
  columns=fullflexible,
  showstringspaces=false,
  breaklines=true,
  numbers=left,
  numberstyle=\tiny\color{gray},
  numbersep=6pt,
  frame=single,
  framerule=0.4pt,
  framesep=2pt,
  xleftmargin=0pt,
  xrightmargin=0pt,
  backgroundcolor=\color{black!3},
  literate=
   *{0}{{{\color{blue}0}}}{1}
    {1}{{{\color{blue}1}}}{1}
    {2}{{{\color{blue}2}}}{1}
    {3}{{{\color{blue}3}}}{1}
    {4}{{{\color{blue}4}}}{1}
    {5}{{{\color{blue}5}}}{1}
    {6}{{{\color{blue}6}}}{1}
    {7}{{{\color{blue}7}}}{1}
    {8}{{{\color{blue}8}}}{1}
    {9}{{{\color{blue}9}}}{1}
    {:}{{{\color{red}:}}}{1}
    {,}{{{\color{red},}}}{1}
    {\{}{{{\color{orange}\{}}}{1}
    {\}}{{{\color{orange}\}}}}{1}
    {[}{{{\color{orange}[}}}{1}
    {]}{{{\color{orange}]}}}{1},
}

\lstdefinestyle{jsoncompact}{
  language=json,
  basicstyle=\ttfamily\footnotesize, 
  aboveskip=3pt,
  belowskip=3pt,
  captionpos=b,
  abovecaptionskip=2pt,
  belowcaptionskip=0pt,
}

\definecolor{custom-gray}{cmyk}{0, 0, 0, 0.7, 1.00}

\begin{document}
\bstctlcite{BSTcontrol}

\title{Human-AI Synergy in Agentic Code Review}


\author{
    Suzhen Zhong,
    Shayan Noei,
    Ying Zou,~\IEEEmembership{Senior Member,~IEEE},
    Bram Adams,~\IEEEmembership{Senior Member,~IEEE}
    \thanks{Suzhen Zhong, Shayan Noei, and Ying Zou are with the Department of Electrical and Computer Engineering, Queen’s University, Kingston, ON K7L 3N6, Canada. E-mail: \{suzhen.zhong, s.noei, ying.zou\}@queensu.ca.}
    \thanks{Bram Adams is with the Maintenance, Construction and Intelligence of Software Lab (MCIS), School of Computing, Queen’s University, Kingston, ON K7L 3N6, Canada. E-mail: bram.adams@queensu.ca.}
}



\maketitle
\begin{abstract}
Code review is a critical software engineering practice where developers review code changes before integration to ensure code quality, detect defects, and improve maintainability. In recent years, AI agents that can understand code context, plan review actions, and interact with development environments have been increasingly integrated into the code review process. However, limited empirical evidence compares the feedback characteristics, interaction patterns, and suggestion outcomes of AI agents and human reviewers. To address this gap, we conduct a large-scale empirical analysis of 278,790 code review conversations across 300 open-source GitHub projects. In our study, we aim to compare the feedback differences provided by human reviewers and AI agents. We investigate how human-AI interaction patterns are associated with subsequent code changes. Moreover, we analyze the adoption of code suggestions provided by human reviewers and AI agents into the codebase and how adopted suggestions change code metrics. We find that over 95\% of AI agent comments focus on Code Improvement or Defect Detection, whereas human reviewers provide a broader range of feedback, including Understanding, Testing, and Knowledge Transfer. Human reviewers exchange 11.8\% more rounds when reviewing AI-generated code than human-written code. Moreover, code suggestions made by AI agents are adopted into the codebase at a significantly lower rate than suggestions proposed by human reviewers (16.6\% vs. 56.5\%). When adopted, suggestions provided by AI agents are associated with significantly larger increases in code complexity and code size than suggestions provided by human reviewers. Our findings suggest that while AI agents can scale defect screening, human oversight remains critical for ensuring suggestion quality and providing contextual feedback that AI agents lack.
\end{abstract}

\newcommand{\rqone}{What are the similarities and differences between the review comments by AI agents and human reviewers?}
\newcommand{\rqtwo}{How do interaction patterns differ between human and AI agent code reviews?}
\newcommand{\rqthree}{How does code quality change after suggestions from human reviewers and AI agents are adopted?} 

\newcommand{\motivation}{\textbf{\textit{Motivation. }}}
\newcommand{\approach}{\textbf{\textit{Approach. }}}
\newcommand{\findings}{\textbf{\textit{Findings. }}}


\section{Introduction}\label{sec:introduction}

With the widespread adoption of generative AI, developers increasingly use AI-powered tools to generate code, expanding codebases at unprecedented scale. GitHub reports nearly 1 billion commits in 2025 with a 178\% surge in generative AI projects~\cite{github_octoverse}, while Google and Microsoft report that AI-generated code now comprises approximately 30\% of new code at their companies~\cite{google_ceo_ai_2024, microsoft_ceo_ai_2024}. However, the quality of AI-generated code remains uncertain and requires validation before integration. Code review, where reviewers examine code changes, provide feedback, and discuss with authors and other reviewers before approving integration, serves as a critical quality checker to ensure quality, detect defects, and improve maintainability. With the increased usage of AI coding assistants in recent years, code production has accelerated beyond the capacity of reviewers; therefore, software teams face increasing pressure to maintain quality standards~\cite{hassan2025agentic}.

\begin{figure}[!t]
    \centering
    \includegraphics[width=\columnwidth]{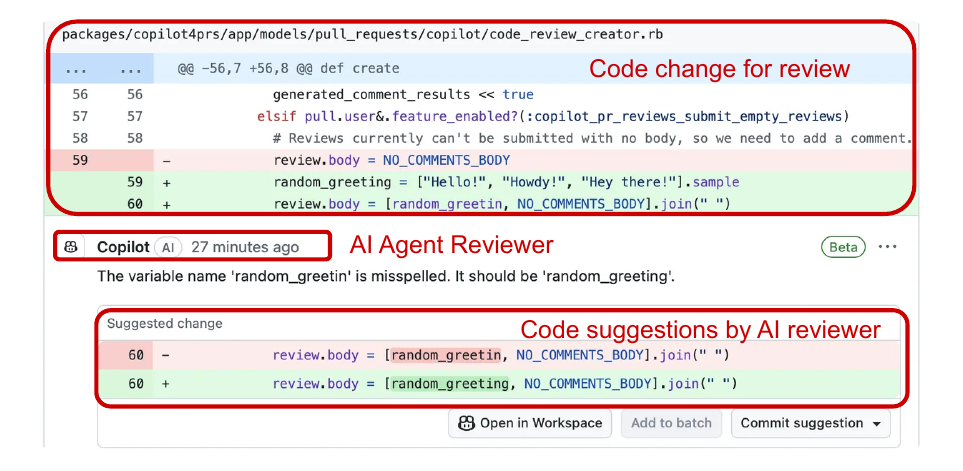}
    \caption{Example of GitHub Copilot agent reviewing a hunk. 
    }
    \label{fig:codereview_example}
    \vspace{-15pt}
\end{figure}

To address the growing gap between code volume and human review capacity, AI agents have been increasingly integrated into the code review process~\cite{coderabbit, github_copilot}. AI agents can understand code context, reason about code logic, and interact with development environments to provide feedback and propose code modifications. Prior work has conducted extensive studies on the capabilities of AI agents, such as defect detection~\cite{10378848} and generating comments on code changes~\cite{li2025riseaiteammatessoftware}. However, prior work lacks understanding of whether feedback from AI agents performs similarly to or differs from feedback by human reviewers, if there exist human-AI collaboration patterns associated with subsequent code changes, and how code metrics change following adoption of AI-agent suggestions.

In this study, we conduct a large-scale empirical analysis of 278,790 code review conversations from 300 open-source GitHub projects, involving human reviewers and AI agents. Figure~\ref{fig:codereview_example} illustrates an inline code review conversation, where a reviewer provides feedback on a code hunk (a block of changed code lines) with a natural language explanation and a proposed code modification. We focus on inline code review conversations because inline comments are attached to a specific block of changed code, called a code hunk, so reviewers provide their feedback at the exact code changes they are discussing. 
To this end, we compare the feedback characteristics provided by AI agents and human reviewers, examine human-AI collaboration patterns by analyzing review conversations, and understand the code quality of code modifications proposed by human reviewers and AI agents.

Our findings provide actionable guidelines for practitioners to assign review tasks based on the strengths of AI agents and human reviewers, streamline review processes to maximize the benefits of human-AI collaboration, and understand where code modifications by AI agents are most effective. We aim to answer the following research questions:

\textbf{RQ1. \rqone} To help practitioners leverage the respective strengths of AI agents and human reviewers in the assignment of review tasks, we compare review feedback by AI agents and human reviewers. We find that over 95\% of AI agent comments focus on \textit{Code Improvement} and \textit{Defect Detection}, while human reviewers provide additional feedback types, such as \textit{Understanding} and \textit{Knowledge Transfer}. In addition, human reviewers write significantly more readable comments, while AI agent reviewers more frequently combine an evaluation of a code weakness with a proposed fix in the same comment.

\textbf{RQ2. \rqtwo} We analyze the interaction sequences within the review conversations to examine how the human-AI interactions relate to whether the reviewed code is subsequently modified. Our analysis shows that human reviewers exchange 11.8\% more rounds when reviewing AI-generated code than human-written code, while 85–87\% of AI agent-initiated reviews end after the first comment without follow-up discussions. Review conversations in which human reviews human-written code are more frequently followed by related code changes than conversations involving AI agents.

\textbf{RQ3. \rqthree} To guide practitioners on when to adopt AI agent suggestions, we compare how frequently suggestions from AI agents and human reviewers are adopted and examine the changes in selected code metrics following their adoption. Our findings illustrate that although AI agents generate more code suggestions, suggestions from human reviewers have a significantly higher adoption rate than suggestions from AI agents~(56.5\% vs. 16.6\%). When adopted, suggestions from AI agents are associated with significantly larger increases in code complexity and code size than suggestions from human reviewers. 

Our work makes the following main contributions:

\begin{itemize}
    \item We present a dataset of 278,790 inline code review conversations from 300 mature open-source GitHub projects from 2022 to November 2025. Our dataset captures various characteristics of code review patterns carried out by human reviewers and AI agents. The dataset allows us to gain insights on how AI agents and human reviewers integrate into collaborative code review workflows.
    \item We characterize the similarities and differences in feedback between AI agents and human reviewers, allowing human practitioners to effectively assign AI agents and human reviewers to appropriate review tasks based on their demonstrated capabilities.
    \item We examine how the suggestions provided by AI agents are adopted into the codebase and the code-metric changes associated with the adopted suggestions, providing evidence to inform practitioners’ decisions about adopting AI agent suggestions.
\end{itemize}
\section{Experiment Setup}\label{sec:methodology}
This section details the experimental setup of our study, covering our methods for data collection, pre-processing, and the analysis approaches for each research question.

\begin{figure*}[t]
    \centering
    \includegraphics[width=\textwidth]{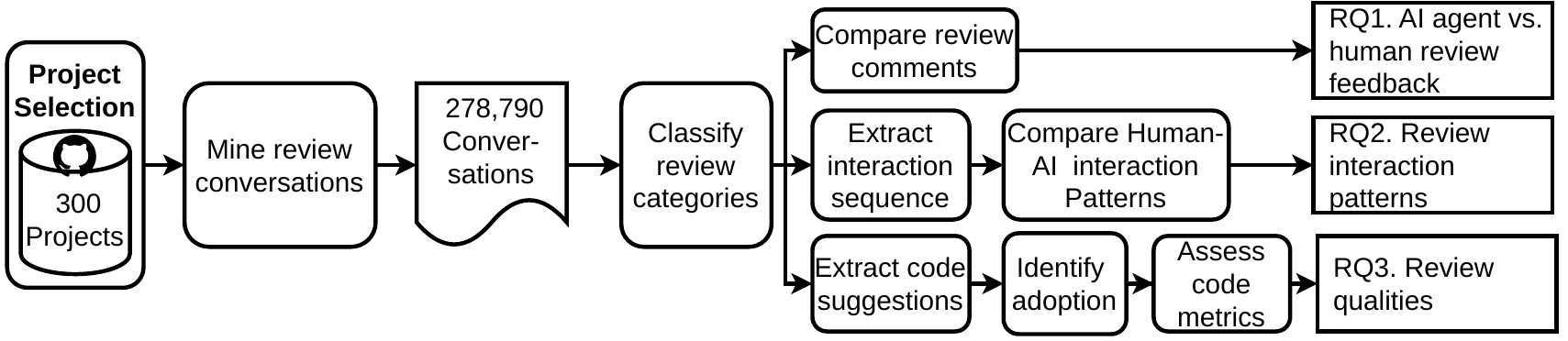}
    \caption{Overview of our approach.}
    \vspace{-10pt}
    \label{fig:approach}
\end{figure*}

\subsection{Overview of Our Approach}\label{sec:overview}
An overview of our approach is shown in Figure~\ref{fig:approach}. First, we systematically select 300 GitHub projects and mine review conversations from closed pull requests. Then, we classify the review categories according to the identity of the authors and reviewers, such as human or AI agent. We then compare review comments by measuring comment-quality attributes and different types of feedback to answer RQ1. For RQ2, we measure the review effort, interaction sequences, and state transitions to examine human-AI collaboration patterns. Finally, we analyze changes in code quality before and after adopted suggestions are applied.

\subsection{Data Collection}
\textbf{Project Selection.}\label{sec:select_projects}
We apply a systematic approach to select the projects for our study. To this end, using the GitHub advance search~\cite{github_search}, we select GitHub projects based on the following criteria:
\begin{itemize}
    \item at least 100 stars~\cite{unmaintain_project_github}, to ensure projects have sufficient community adoption and maintenance activity;
    \item at least one closed pull request per month from 2022 to November 2025, ensuring projects maintain consistent review activity throughout the study period; and
    \item at least 100 pull requests reviewed by AI agents
\end{itemize}

Our applied filtering ensures that the selected projects are actively maintained, have sufficient review history, and contain enough AI agent reviews for meaningful analysis, yielding a final sample of 300 projects and 54,330 closed PRs.

We use the GitHub API~\cite{github_rest_api} to identify whether pull requests are reviewed by AI agents. However, the GitHub API distinguishes reviewer account types as \textit{Human User} or \textit{Bot}, without specifying whether a bot is AI-based. Bots on GitHub include a wide range of automated tools, such as GitHub Actions bot~\cite{github_actions_bot}, which are not AI-based code reviewers. To accurately identify AI agent reviewers, for each bot account name, we manually search on their official website and documentation to determine whether the bot is AI-based. This process identified 16 AI-based code review bots, which are included in our replication package~\cite{aicodereview2026}. For these bots, we further search official blogs and changelogs to identify when agentic capabilities were introduced. Agentic capabilities refer to the ability to reason about code, plan review actions, and interact with the development environment~\cite{ibm_ai_agents}. We classify reviews as AI agent reviews only from the agent capability announcement date onward. For instance, GitHub Copilot reviews are classified as AI agent reviews from May 19, 2025~\cite{CopilotAgent2025}.


\textbf{Mining Review Conversations.} 
Each PR contains PR-level conversations and inline conversations. PR-level conversations discuss the entire pull request, while inline conversations focus on specific code hunks (see Figure~\ref{fig:codereview_example}). We collect inline conversations because the direct linkage to code hunks enables tracing feedback to specific code changes. Since each PR contains multiple code hunks, one PR yields multiple inline conversations. For each conversation, we record repository name, PR ID, PR author, and timestamps; for each comment within a conversation, we record reviewer identity, content, and timestamp.

\begin{table}[!t]
\centering
\caption{Distribution of inline code review conversations across four review categories based on reviewer and code author identities.}
\label{tab:dataset}
\begin{adjustbox}{max width=\linewidth}
\begin{tabular}{lrr}
\toprule
\textbf{Category} & \textbf{Conversations} & \textbf{\%} \\
\midrule
\textbf{Human reviews} (Total) & \textbf{123,393} & \textbf{44.3\%} \\
\textbf{HRH} (Human reviews Human-written code) & 116,874 & 41.9\% \\
\textbf{HRA} (Human reviews Agent-generated code) & 6,519 & 2.3\% \\
\midrule
\textbf{Agent reviews} (Total) & \textbf{155,397} & \textbf{55.7\%} \\
\textbf{ARH} (Agent reviews Human-written code) & 154,469 & 55.4\% \\
\textbf{ARA}\textsuperscript{*} (Agent reviews Agent-generated code) & 928 & 0.3\% \\
\midrule
\textbf{Total} & \textbf{278,790} & \textbf{100\%} \\
\bottomrule
\end{tabular}
\end{adjustbox}
\vspace{0.25em}
\raggedright\footnotesize\textsuperscript{*}In ARA, 94.0\% (872) involve the same agent reviewing its own code, while 6.0\% involve different agents (e.g., CodeRabbit reviews Copilot's code).
\vspace{-15pt}
\end{table}

\textbf{Classifying Review Categories.} We classify inline code review conversations into four review categories based on the identity of the PR author and the first reviewer who comments on each code change.
For each conversation, we extract the PR author from the PR metadata and the first commenter from the conversation. Using the AI agent identification approach (described in Section~\ref{sec:select_projects}), we determine whether each account is a human or an AI agent. The PR author identity determines whether the code is human-written or agent-generated. The first commenter's identity determines whether a human or an AI agent initiated the review. For example, the PR author \textit{copilot[bot]} and the first commenter \textit{copilot[bot]} are both AI agents, classifying the conversation as \textit{AI agent reviews agent-generated code}. This classification yields four review categories: 
\begin{itemize}
    \item Human reviews human-written code (HRH)
    \item Human reviews agent-generated code (HRA)
    \item AI agent reviews human-written code (ARH)
    \item AI agent reviews agent-generated code (ARA)
\end{itemize} 

Table~\ref{tab:dataset} summarizes the distribution across the four review categories. Human reviews (i.e., HRH and HRA) account for 44.3\% of conversations, including 41.9\% HRH and 2.3\% HRA. AI-agent reviews (i.e., ARH and ARA) account for 55.7\% of conversations, including 55.4\% ARH and 0.3\% ARA. 

\subsection{Labeling Conversations}\label{sec:llm_labeling}
To understand whether AI agents and human reviewers focus on similar or different aspects of code review, we adopt the taxonomy of Bacchelli and Bird~\cite{6606617}, who identify nine feedback categories that capture the actual results of code review, including \textit{Code Improvement}, \textit{Defect Detection}, \textit{External Impact}, \textit{Knowledge Transfer}, \textit{Misc}, \textit{No Feedback}, \textit{Social}, \textit{Testing}, and \textit{Understanding}. The detailed description of each feedback type is specified in Table~\ref{tab:feedback_type_taxonomy}.

\begin{table}[t]
\centering
\caption{Feedback types from Bacchelli and Bird~\cite{6606617} with rephrased descriptions.}
\label{tab:feedback_type_taxonomy}
\begin{tabularx}{\columnwidth}{@{}l>{\raggedright\arraybackslash}X@{}}
\toprule
\textbf{Feedback Type} & \textbf{Description} \\
\midrule
Code Improvement & Suggestions to enhance code clarity, style, structure, or maintainability without fixing defects. \\
\midrule
Defect Detection & Identification of functional, logical, or correctness issues in the proposed changes. \\
\midrule
External Impact & Comments about broader system-level consequences beyond the local code diff. \\
\midrule
Knowledge Transfer & Reviewer explains concepts, conventions, best practices, or provides learning resources. \\
\midrule
Misc & Comments that do not fit any defined category or are context-irrelevant. \\
\midrule
No Feedback & Conversations where the reviewer provides no substantive technical feedback. \\
\midrule
Social & Interpersonal statements not directly tied to technical content (e.g., appreciation, encouragement). \\
\midrule
Testing & Comments about adding, updating, or fixing tests and test coverage. \\
\midrule
Understanding & Clarification questions to understand context, rationale, design decisions, or implementation intent. \\
\bottomrule
\end{tabularx}
\vspace{-15pt}
\end{table}


Due to the large scale of our dataset with 278,790 conversations, manually labeling them into one of nine feedback types is infeasible. Therefore, we adopt an LLM-based annotation approach to classify review comments into feedback types, following Ahmed et al.~\cite{ahmed2025can}, who demonstrate that large language models such as Claude-3.5-Sonnet~\cite{anthropic_claude35sonnet} and GPT-4~\cite{openai_gpt4} achieve human-level accuracy in software engineering classification tasks. Figure~\ref{fig:feedback_prompt} shows the prompt that instructs the model to classify each review comment into one type of feedback. In addition to the feedback-type label, GPT-4.1-mini returns a confidence score ranging from 1 to 10 for each classification. Overall, 96.4\% of the 278,790 review conversations receive confidence scores of 8 or higher. Following prior work~\cite{gligoric-etal-2025-unconfident,Confidence_GPTlabel}, which reported approximately 81\% agreement between GPT-4o and human labels on a text-classification task when confidence scores exceeded 0.8 on a scale of 0 to 1, we treat confidence scores below 8 as low confidence and exclude those conversations from the feedback-type analysis. To validate reliability, the first author manually labels a statistically representative sample of 383 comments with 95\% confidence level and 5\% margin of error~\cite{ahmad2017determining}. Comparing manual labels against LLM classifications yields Cohen's $\kappa$~\cite{cohen1960coefficient} of 0.85, indicating almost perfect agreement between human evaluator and GPT-4.1-mini feedback.

\begin{figure}[t]
\centering
\small
\fbox{\parbox{0.95\columnwidth}{
\textbf{Prompt:} Based on the feedback type definitions from Table~\ref{tab:feedback_type_taxonomy}, classify the following code review comment into the most appropriate feedback type. Return the feedback type with a confidence score (1-10). Code review comment: \texttt{\{comment\_body\}}
}}
\caption{Prompt for labeling feedback types.}
\label{fig:feedback_prompt}
\vspace{-15pt}
\end{figure}

\subsection{Interaction Pattern Extraction}\label{sec:interaction_pattern_extraction} 
To capture the detailed flow of interactions between humans and AI agents, we extract the sequence of comment authors from each conversation. Each sequence starts with human-written code (HC) or agent-generated code (AC) based on the pull request author identity. Each comment author is classified as a human or AI agent using the identification approach described in Section~\ref{sec:select_projects}. Following prior work~\cite{rahman2017predicting,sun2025doesaicodereview}, we classify the outcome of a review conversation as ``With related code changes'' when subsequent changes address at least one review comment and are retained in the main codebase when the pull request is merged. Otherwise, we classify the review outcome as ``Without related code changes''. For example, if a reviewer requests a null check before a variable is accessed, and a subsequent commit adds that check and retains it in the merged code, we classify the conversation as ``With related code changes''. If the check is not added, is reverted before merging, or belongs to a pull request that is closed without merging, we classify the conversation as ``Without related code changes''.

In each inline human-AI review conversation, GitHub provides the range of lines for the reviewed code, the conversation timestamp, subsequent commit changes, and the final pull request status (i.e., merge or reject). We collect code changes made to the reviewed file during the review conversation and before the pull request is closed. A review conversation is classified as ``Without related code changes'' when there is no subsequent commit that modifies the reviewed code or when the resulting modifications are not integrated into the codebase. This occurs if the modification is removed before merging or if the pull request is closed without merging. For comments containing code snippets proposed as replacements, we use Jaccard similarity~\cite{codeclone} to compare the committed replacement with both the suggested snippet and the original reviewed code. We consider the change associated with the suggestion when its similarity to the suggested snippet is higher than its similarity to the original code. For comments containing only natural-language suggestions, we follow the LLM-based approach proposed by Sun et al.~\cite{sun2025doesaicodereview}. We provide the LLM with the review comment, the reviewed file path, and the relevant reviewed-code diff to determine whether the subsequent changes address the comment. We manually validate the LLM's assessments on a sample of 50 cases and find that all 50 agree with our manual labels. To this end, we only classify a conversation as ``With related code changes'' when either the similarity comparison or the LLM assessment indicates that the comment is addressed and the code changes are retained in the merged code.

\subsection{Code Metric Assessment}\label{sec:assess_code_metrics}
To examine changes in code metrics following suggestion adoption, we measure properties such as code size, complexity, coupling, and cohesion before and after applying each adopted suggestion. To measure these structural properties, we utilize SciTools Understand~\cite{scitools_understand_metrics}, a static analysis tool widely recognized in software engineering research~\cite{alikhashashneh2018ml, jin2021ci, relaesewise}. We compute 111 code metrics on the source file before and after applying each of the 8,985 adopted suggestions, such as lines of code, executable statements, complexity, coupling, and cohesion~(the full list is provided in replication package~\cite{aicodereview2026}).

\subsection{Cross-Agent Analysis}\label{sec:cross-agent-analysis}
To examine whether AI agent reviewers developed by different vendors vary in their review behavior, we separately analyze agents that generate at least 383 review comments or code suggestions. This criterion selects CodeRabbit~\cite{coderabbit}, GitHub Copilot~\cite{github_copilot}, Gemini Code Assist~\cite{gemini_code_assist}, Cursor~\cite{cursor_web}, Claude Code~\cite{claude_code}, Sourcery AI~\cite{sourcery_web}, and Greptile~\cite{Greptile_web} for all three RQs. To ensure comparable sample sizes, we randomly sample 383 observations per agent, corresponding to a 95\% confidence level and a 5\% margin of error~\cite{ahmad2017determining}.
\section{Results}\label{sec:results}

\subsection{RQ1: \rqone}\label{sec:RQ1}

\motivation
Prior work examines AI-generated code review feedback~\cite{sun2025doesaicodereview,olewicki2024impactllmbasedreviewcomment}, without comparing how AI agent and human reviewer feedback differ in content and focus. Understanding how AI agent and human reviewer feedback differ in content and technical focus is critical to determining the review tasks that AI agents can handle or require human participation. Such information can help practitioners assign review tasks based on the respective strengths of AI agents and human reviewers, and identify feedback types where human oversight remains essential. In this RQ, we compare the feedback types and comment quality of AI agents and human reviewers to establish a baseline for understanding how human reviewers and AI agents collaborate in the code review process.

\approach
To compare AI-agent and human-reviewer comments, we classify their feedback types to identify differences in review focus. We also assess comment readability and characterize comment content using quality attributes.

\textbf{Classifying feedback types.}
To understand whether AI agents and human reviewers focus on similar or different aspects of code review, we classify their feedback types using the taxonomy of Bacchelli and Bird~\cite{6606617}. This taxonomy identifies nine feedback categories capturing actual outcomes of code review (see Table~\ref{tab:feedback_type_taxonomy}). Using the automated labeling approach described in Section~\ref{sec:llm_labeling}, we classify the first comment of each conversation into one of the nine feedback categories, as the first comment reflects the reviewer's independent assessment before any discussion begins. We then compare the distribution of feedback types across the four review categories, namely HRH, HRA, ARH, and ARA. 

\textbf{Measuring comment quality.} To examine the quality of review comments written by human and AI agent reviewers, we follow prior work that measures the readability of a comment~\cite{rahman2017predicting} and its quality attributes~\cite{yang2023evacrc}, as listed in Table~\ref{tab:rq1_comment_quality}.

\begin{table}[t]
\centering
\renewcommand{\arraystretch}{1.08}
\caption{Comment quality metrics from~\cite{rahman2017predicting} and~\cite{yang2023evacrc}.}
\label{tab:rq1_comment_quality}
\begin{tabularx}{\columnwidth}{
    @{}
    >{\raggedright\arraybackslash}p{0.15\columnwidth}
    @{\hspace{6pt}}
    >{\raggedright\arraybackslash}X
    >{\centering\arraybackslash}p{0.10\columnwidth}
    @{}
}
\toprule
\textbf{Metric} & \textbf{Description} & \textbf{Study} \\
\midrule

Reading Ease
& Flesch-Kincaid readability score, higher values indicate text that is easier to read.
& \cite{rahman2017predicting} \\
\cmidrule(l){1-3}

Emotion
& Does it comment in a kind way?
& \cite{yang2023evacrc} \\
\cmidrule(l){1-3}

Question
& Does it ask questions or ask for confirmation?
& \cite{yang2023evacrc} \\
\cmidrule(l){1-3}

Evaluation
& Does it present evaluation or show code weakness?
& \cite{yang2023evacrc} \\
\cmidrule(l){1-3}

Suggestion
& Does it provide suggestions for improvements?
& \cite{yang2023evacrc} \\

\bottomrule
\end{tabularx}
\vspace{-15pt}
\end{table}

\begin{itemize}

\item \textbf{Reading ease.} To assess the readability of comments, we use Flesch-Kincaid readability tests~\cite{rahman2017predicting} to measure the complexity of the comment text. The scores of Flesch-Kincaid range from 0 to 100. Higher scores indicate greater readability.

\item \textbf{Quality attribute.} We classify each comment using four binary attributes, namely Emotion, Question, Evaluation, and Suggestion~\cite{yang2023evacrc}, as shown in Table~\ref{tab:rq1_comment_quality}. We use GPT-4.1-mini with the attribute definitions and few-shot examples from the same study~\cite{yang2023evacrc} (see Figure~\ref{fig:qa_prompt}). To validate the automated labels, the first author independently annotates a sample of 383 review conversations. The resulting Cohen's $\kappa$ of 0.9 indicates almost perfect agreement with the automatic labels~\cite{cohen1960coefficient}.

\end{itemize}

\begin{figure}[h]
\centering
\small
\fbox{\parbox{0.95\columnwidth}{
\textbf{Prompt:} Based on the four quality attribute definitions from Table~\ref{tab:rq1_comment_quality} (Emotion, Question, Evaluation, Suggestion), classify the following code review comment by answering each attribute with 1 (Yes) or 0 (No). Examples are provided in \texttt{\{example\}}. Code review comment: \texttt{\{comment\_body\}}
}}
\caption{Prompt for labeling quality attribute.}
\vspace{-5pt}
\label{fig:qa_prompt}
\end{figure}


To compare the four review categories (i.e., HRH, HRA, ARA, ARH), we rank the four review categories by reading ease using the Scott-Knott ESD test~\cite{Tantithamthavorn2017}, which statistically clusters categories that do not significantly differ from one another and ranks the resulting clusters by magnitude. We then use a chi-square test~\cite{Hahn01101984} to determine whether the distributions of Emotion, Question, Evaluation, and Suggestion differ across the four review categories.

\begin{figure}[!t]
    \centering
    \begin{minipage}{\columnwidth}\centering
        \includegraphics[width=\columnwidth]{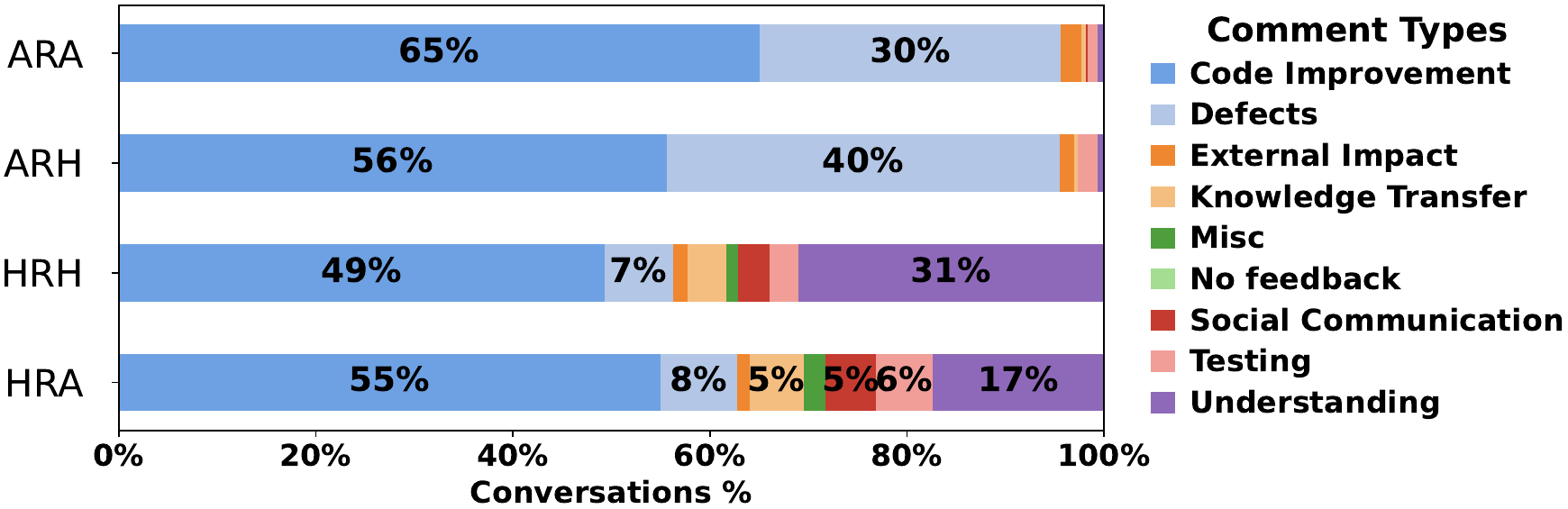}\\
        \scriptsize (a) Review feedback categories~\cite{6606617}.
    \end{minipage}%
    \vspace{5pt}
    
    \begin{minipage}{\columnwidth}\centering
        \includegraphics[width=\columnwidth]{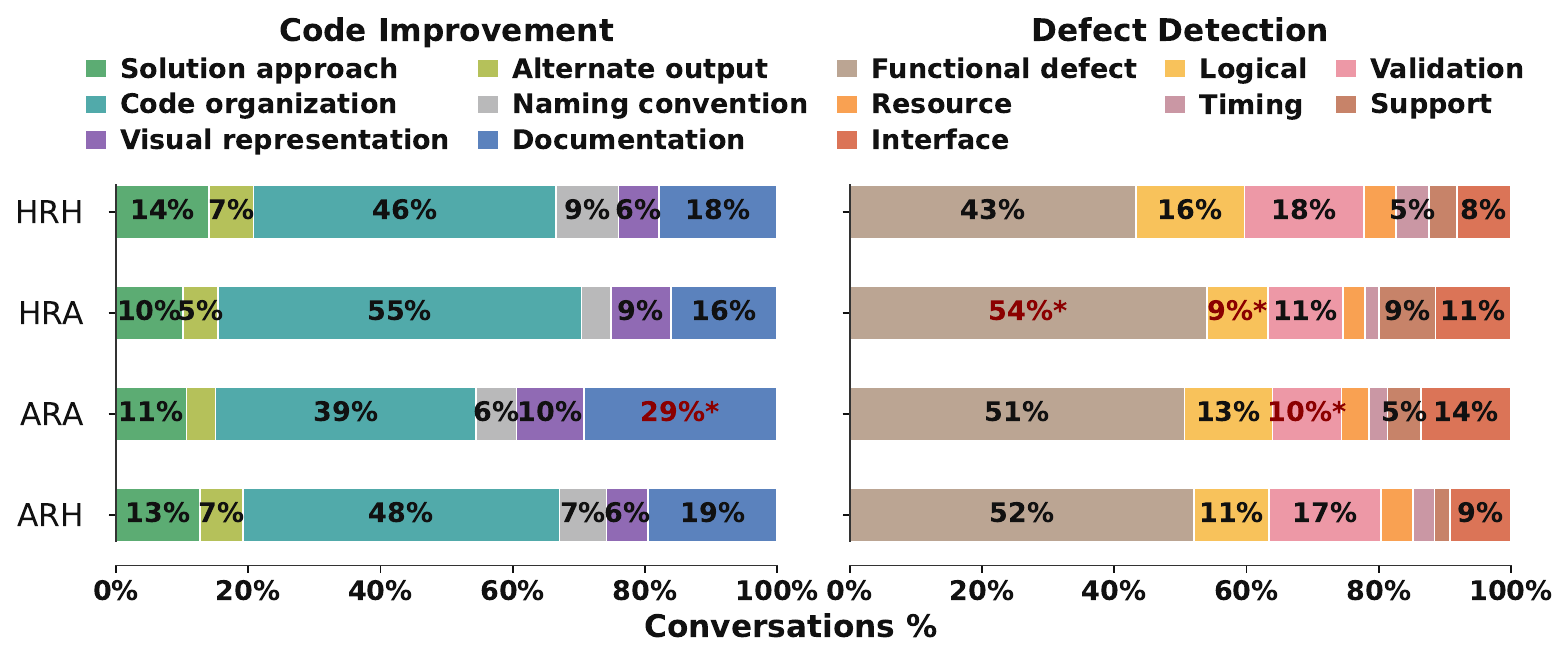}\\
        \scriptsize (b) Defect Detection and Code Improvement subcategories~\cite{Turzo2023}. $*$ indicates a statistically significant difference from HRH with a small effect size.
    \end{minipage}
    \caption{Distribution of feedback categories across the four review categories.} 
    \label{fig:rq1_feedback}
    \vspace{-10pt}
\end{figure}


\findings \textbf{AI agent reviews focus on Code Improvement and Defect Detection, while human reviewers provide a wider diversity of feedback.} As shown in Fig.~\ref{fig:rq1_feedback}, \textit{Code Improvement} and \textit{Defect Detection} include over 95\% of AI agent comments. Human reviewers also focus on \textit{Code Improvement} at 49--55\%, but provide more diverse feedback types, such as \textit{Understanding}, \textit{Testing}, and \textit{Knowledge Transfer}. To better understand the technical issues related to Code Improvement and Defect Detection, we refine our analysis of \textit{Code Improvement} and \textit{Defect Detection} comments using the fine-grained categories from Turzo and Bosu~\cite{Turzo2023}. As shown in Figure~\ref{fig:rq1_feedback}, AI agents provide more documentation feedback when reviewing agent-generated code than human reviewers provide when reviewing human-written code, with 29\% versus 18\%. For \textit{Defect Detection}, human reviewers report more functional defects but fewer logical defects when reviewing agent-generated code than human-written code, with 54\% versus 43\% and 9\% versus 16\%, respectively. AI agents report fewer validation issues when reviewing agent-generated code than human reviewers report when reviewing human-written code, with 10\% versus 18\%. However, all four statistically significant differences have small effect sizes, indicating that the differences in focus within code improvement and defect detection are limited in magnitude.

\textbf{Human reviewers frequently provide \textit{Understanding} feedback, accounting for 31\% of comments when reviewing human-written code and 17\% when reviewing agent-generated code.} For example, a human reviewer asks ``Why are we removing the MockTokenCredential in playback?'' before providing other suggestions. Human reviewers also provide \textit{Knowledge Transfer} feedback, accounting for 4--6\% of comments when reviewing both human-written and agent-generated code, by pointing to existing code, such as ``Refer to sdk/hdinsight/ci.yml for batch release setup.'' These results indicate that human reviewers more frequently clarify design intent through questions and share project-specific references, whereas AI agents less frequently provide comments with references to support their proposed fixes.

\begin{figure}[!t]
    \centering
    \includegraphics[width=\columnwidth]{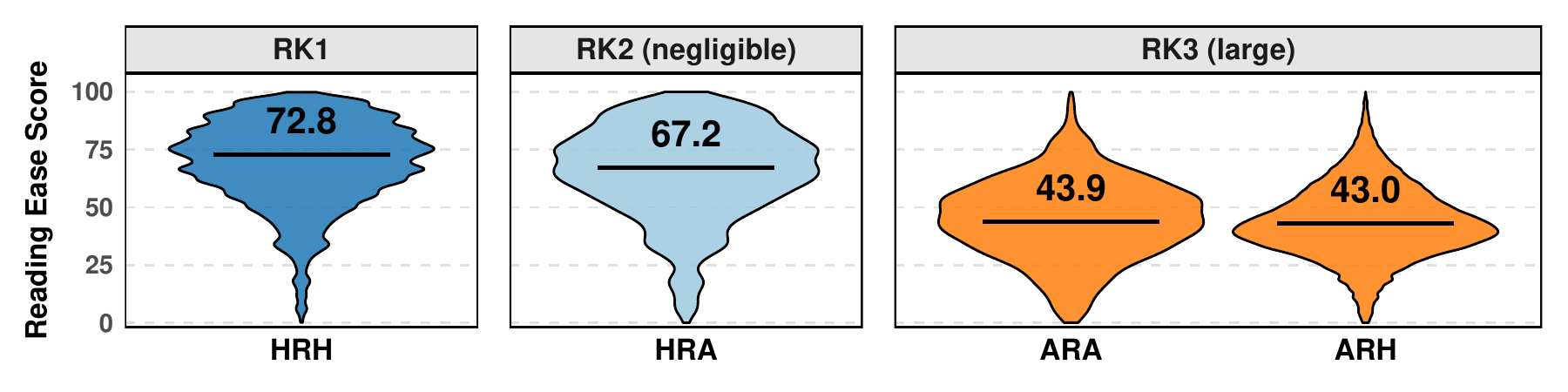}
    \caption{Reading ease score ranking (Scott-Knott ESD) across four review categories.}
    \label{fig:rq1_readease}
\end{figure}

\textbf{Human reviewers write significantly more readable comments than AI agent reviewers, regardless of whether they review human-written or agent-generated code.} As shown in Figure~\ref{fig:rq1_readease}, human reviewers have average reading-ease scores of 67.2–72.8, compared with 43.0–43.9 for AI agent reviewers, with large effect sizes. Table~\ref{tab:attribute_combinations} further shows that AI agent reviewers are significantly more likely than human reviewers to identify a code weakness (Evaluation) and suggest a correction without asking a question, with 87.7\% versus 31.3\% for human-written code and 94.5\% versus 31.9\% for agent-generated code. Both differences have large effect sizes. Human reviewers more frequently provide a suggestion without either asking a question or identifying a code weakness when reviewing agent-generated code than when reviewing human-written code (32.6\% versus 11.7\%, with a medium effect size). These findings suggest that AI agent reviewers typically present a code weakness and its proposed fix directly within the same comment, while human reviewers more regularly ask questions to seek clarification.

\begin{table}[t]
\centering
\caption{Distribution of quality attribute combinations across review categories. Emo=Emotion, Que=Question, Eva=Evaluation, Sug=Suggestion.}
\label{tab:attribute_combinations}
\small
\setlength{\tabcolsep}{3pt}
\begin{tabularx}{\columnwidth}{*{4}{c} *{4}{>{\centering\arraybackslash}X}}
\toprule
Emo & Que & Eva & Sug & HRH & HRA & ARA & ARH \\
\midrule
1 & 1 & 1 & 1 & 8.9\% & 4.7\% & 1.6\%\textsuperscript{*} & 4.2\% \\
1 & 1 & 1 & 0 & 4.7\% & 2.3\% & -- & -- \\
1 & 1 & 0 & 1 & 8.1\% & 5.2\% & -- & -- \\
1 & 1 & 0 & 0 & 21.1\% & 10.4\%\textsuperscript{*} & -- & -- \\
1 & 0 & 1 & 1 & 31.3\% & 31.9\% & \textcolor{red}{94.5\%\textsuperscript{***}} & \textcolor{red}{87.7\%\textsuperscript{***}} \\
1 & 0 & 1 & 0 & 8.6\% & 11.0\% & 2.3\%\textsuperscript{*} & 6.8\% \\
1 & 0 & 0 & 1 & 11.7\% & \textcolor{red}{32.6\%\textsuperscript{**}} & 1.0\%\textsuperscript{*} & 1.3\%\textsuperscript{*} \\
1 & 0 & 0 & 0 & 5.0\% & 1.6\% & -- & -- \\
\midrule
\multicolumn{4}{c}{Total} & 100\% & 100\% & 100\% & 100\% \\
\bottomrule
\end{tabularx}
\par\vspace{0.25em}\noindent
\raggedright\footnotesize $*$ mark an HRA/ARA/ARH value that significantly differs from HRH with chi-square test. $*/**/***$ shows small/medium/large effect size.
\vspace{-15pt}
\end{table}

\textbf{Cross-agent analysis shows that Code Improvement is the most common feedback type for six agents, whereas Cursor mainly focuses on identifying defects without proposing fixes.} As shown in Figure~A1 in the Appendix~\cite{online_appendix}, Code Improvement is the most frequent feedback type for six agents (49--62\%). Claude Code provides fewer Defect Detection comments than AI agents overall (20\% vs. 40\%), with a small effect size. In contrast, Cursor focuses almost exclusively on Defect Detection (99\% vs. 40\% overall), with a large effect size. For Reading Ease, GitHub Copilot, CodeRabbit, Claude Code, and Greptile show no significant differences from the overall AI-agent results. Gemini Code Assist has a higher average score, while Sourcery AI and Cursor have lower scores, with small effect sizes (see Figure~A2 in the Appendix~\cite{online_appendix}). Moreover, 71.3\% of Greptile comments identify a code weakness (\textit{Evaluation}) and propose an improvement (\textit{Suggestion}), compared with 91.1\% overall, with a medium effect size (see Table~A2 in the Appendix). Cursor comments more often identify a code weakness without proposing an improvement (86.4\% vs. 4.6\% overall) and less often both identify a code weakness and propose an improvement, with large effect sizes. These results indicate that treating AI agent reviewers as a single category generally reflects the behavior of most agents but may not fully capture Cursor's higher concentration of defect-detection comments and lower tendency to suggest solutions.

\smallskip
\begin{Summary}{Summary of RQ1}{
AI agent reviews focus on \textit{Code Improvement} and \textit{Defect Detection}, while human reviewers provide greater diversity with \textit{Understanding}, \textit{Testing}, \textit{Knowledge Transfer}, and \textit{Social Communication}. Human reviewers write significantly more readable comments than AI agent reviewers, while AI agent reviewers more often combine a diagnosed weakness with a proposed fix in the same comment. 
}
\end{Summary}
\subsection{RQ2: \rqtwo}\label{sec:RQ2}

\motivation In RQ1, we find that AI-agent reviews concentrate on code improvement and defect detection, while human reviewers more frequently ask clarification questions. However, these comment-level differences do not show how humans and AI agents interact over multiple review exchanges or whether the reviewed code is subsequently modified. In this RQ, we examine interaction sequences and whether each review conversation is followed by a related code change. 

\approach To characterize collaboration patterns between human reviewers and AI agents, we first extract interaction sequences (Section~\ref{sec:interaction_pattern_extraction}) from each conversation to capture who comments and in what order. We then model the sequences as finite state machines to represent recurring patterns of human and AI agent participation. Finally, we measure review effort and transition probabilities to characterize how interaction patterns differ across review categories.

\textbf{Modeling interaction sequences.} To examine how human and AI-agent participation patterns are associated with subsequent code changes, we model the review process within each review category as a
finite state machine (FSM)~\cite{wagner2006modeling}. FSMs are widely used in process mining in software
engineering to represent sequential processes~\cite{softwareprocessdiscovery,ferreira2007approaching}. An FSM includes states, transitions between states, and terminal conditions. The FSM is suitable for modeling review conversations because conversations have discrete states (who is commenting), defined transitions (comment flow between contributors), and terminal outcomes indicating whether the conversation is followed by a related code change. Following the FSM structure, we define four state types:
\begin{itemize}
\item \textit{H} refers to human comments;
\item \textit{A} represents AI agent comments;
\item \textit{CC} indicates conversation followed by related code changes;
\item \textit{NCC} refers to the conversation not followed by a related code change.
\end{itemize}
With states defined, the state-transition function captures how conversations move from
one contributor state to the next across multiple comment rounds. We define three transition types: (1) self-loops, where the same contributor type comments multiple times consecutively (e.g., H$\rightarrow$H$\rightarrow$H); (2) cross-transitions, where comment authorship switches between human and AI agent (e.g., H$\rightarrow$A); and (3) terminal transitions, where the conversation ends at CC or NCC.

\begin{figure*}[!t]
  \centering
  \begin{minipage}{0.285\textwidth}\centering
    \includegraphics[width=\linewidth]{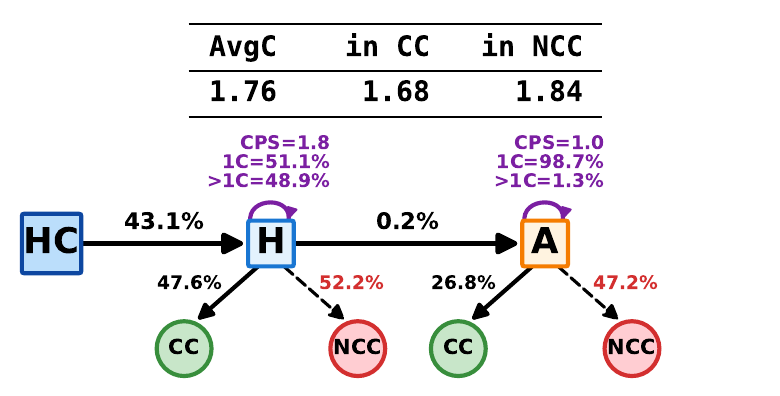}\\
    \scriptsize (a) HRH
  \end{minipage}\hfill
    \begin{minipage}{0.715\textwidth}\centering
    \includegraphics[width=\linewidth]{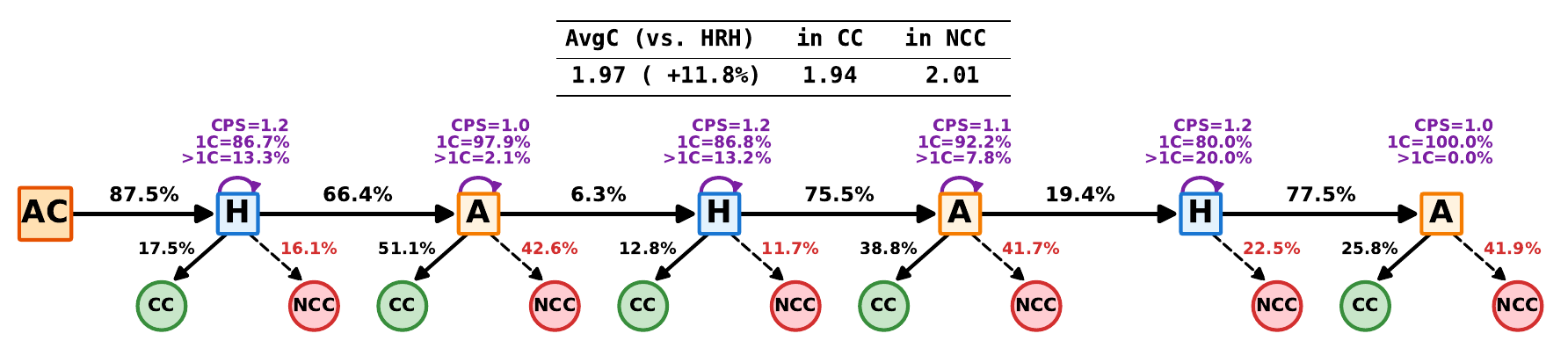}\\
    \scriptsize (b) HRA
  \end{minipage}
  
  \vspace{0.3cm}
  \begin{minipage}{0.415\textwidth}\centering
    \includegraphics[width=\linewidth]{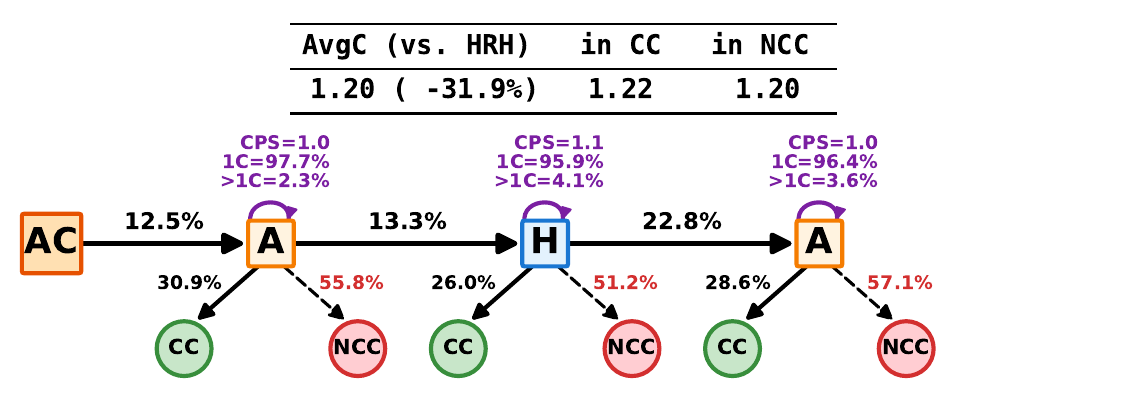}\\
    \scriptsize (c) ARA
  \end{minipage}\hfill
  \begin{minipage}{0.585\textwidth}\centering
    \includegraphics[width=\linewidth]{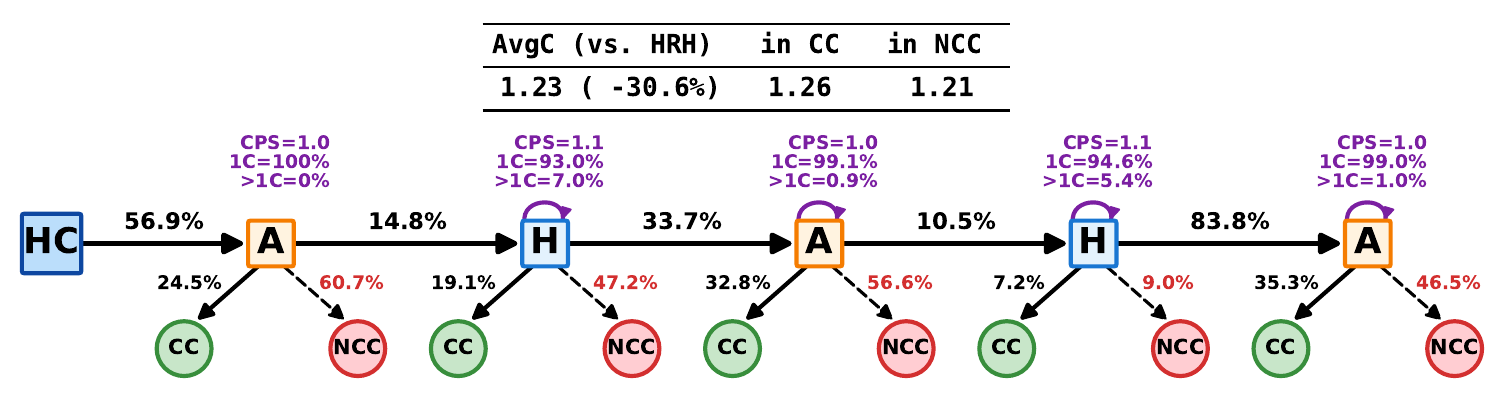}\\
    \scriptsize (d) ARH
  \end{minipage}
  \caption{Interaction patterns modeled as finite state machines (FSMs). Each sub-figure includes a summary table showing the average number of comments per conversation (AvgC) for conversations with and without related code changes and the percentage difference from the HRH baseline. The FSM states represent the contributor type: HC (Human Creator), AC (Agent Creator), H (Human reviewer), and A (AI Agent reviewer), and show whether conversations terminate with related code changes (CC) or without related code changes (NCC). Each state shows how many comment rounds occur at that state on average (CPS), and what percentage of conversations pass through with exactly one comment (1C) or more than one comment (>1C).
  }
  \label{fig:rq2_all}
  \label{fig:rq2_hra}
  \label{fig:rq2_hrh}
  \label{fig:rq2_ara}
  \label{fig:rq2_arh}
  \vspace{-5pt}
\end{figure*}

\textbf{Comparing interaction patterns.}
We aim to compare the interaction patterns within each review category in terms of review
effort, the number of comments that each contributor exchanges consecutively, and the likelihood that a
conversation terminates with or without a related code change at each contributor state. We use HRH as the baseline, as it represents the traditional code review process without AI involvement. For each review category, we construct one FSM from all conversations in the category and measure three aspects of interaction patterns.
\begin{itemize}
\item \textbf{Review effort.} We count the average of comments per conversation and compare each AI-involved category against the HRH baseline. The comparison indicates whether AI participation increases the amount of back-and-forth discussion before the conversation terminates.
\begin{align}
\text{AvgC} &= \frac{\sum_{i=1}^{n} C_i}{n} \\
\text{vs. HRH} &= \frac{\text{AvgC} - \text{AvgC}_{\text{HRH}}}{\text{AvgC}_{\text{HRH}}} \times 100\%
\end{align}
\textit{where $C_i$ is the number of comments in conversation $i$, and $n$ is the total number of conversations in the review category.} 

\item \textbf{Self-loop intensity.} We compute the average number of comments per state and the percentage of conversations with exactly one comment versus more than
one comment, indicating whether a contributor posts multiple consecutive comments
before the conversation switches to the other contributor or terminates.
\begin{align}
\text{CPS} &= \frac{\sum_{j=1}^{m} C_j}{m} \\
\text{1C} &= \frac{\#\text{conversations with 1 comment}}{m} \times 100\% 
\end{align}
\textit{where $C_j$ is the number of comments at the state for conversation $j$, and $m$ is the number of conversations reaching the state.}

\item \textbf{Transition probability.} We compute the probability of a conversation
terminating with a related code change~(CC), terminating without a related code change~(NCC), or continuing to the next contributor state.
\end{itemize}

\begin{figure}[!t]
\centering
\small
\setlength{\fboxsep}{3pt}

\fbox{%
\parbox{\dimexpr\columnwidth-2\fboxsep-2\fboxrule\relax}{%
{\raggedright
\textbf{HRA: A human reviews AI-generated code
(Agoric/agoric-sdk PR~\#11967, 16 turns)}\par}
\colorbox{orange!20}{%
\parbox{\dimexpr\linewidth-2\fboxsep\relax}{%
\raggedright
\texttt{\scriptsize AC: invokeEntry() in ymax-planner/resolver.ts}
}}

\colorbox{blue!10}{%
\parbox{\dimexpr\linewidth-2\fboxsep\relax}{%
\raggedright
\textbf{H$_1$:} Can you split the ymax-planner changes into a
separate PR?
}}

\colorbox{gray!15}{%
\parbox{\dimexpr\linewidth-2\fboxsep\relax}{%
\raggedright
\textbf{A:} I'll revert the ymax-planner changes and create a
separate PR for them.
}}

\colorbox{blue!10}{%
\parbox{\dimexpr\linewidth-2\fboxsep\relax}{%
\raggedright
\textbf{H$_1$:} Your offer included making the follow-up PR.
Where is it?
}}

\colorbox{gray!15}{%
\parbox{\dimexpr\linewidth-2\fboxsep\relax}{%
\raggedright
\textbf{A:} I cannot directly create PRs in GitHub.
}}

\hspace{1.5em}\textit{... 11 more turns ...}

\colorbox{blue!10}{%
\parbox{\dimexpr\linewidth-2\fboxsep\relax}{%
\raggedright
\textbf{H$_1$:} I wish you'd tell me \textit{before} you offer
to do something that you cannot do.
}}
}}

\caption{An example of a Human-Reviews-Agent conversation.}
\label{fig:four_examples}
\vspace{-10pt}
\end{figure}

\textbf{Human reviewers exchange 11.8\% more comments when reviewing AI-generated code than human-written code.} As shown in Figure~\ref{fig:rq2_all}, when humans review human-written code, 99.8\% of conversations terminate after the initial round of human review without AI participation, averaging 1.76 exchange rounds. When humans review AI-generated code, two thirds of conversations extend beyond the initial round of human review, and AI agents continue the conversation with further rounds of human-AI back-and-forth discussion, resulting in 11.8\% longer conversations on average. For example~(see Figure~\ref{fig:four_examples}), the AI agent promises actions it cannot deliver, such as creating pull requests. The human reviewer concludes: ``I wish you'd tell me before you offer to do something that you cannot do.'' These additional exchanges may reflect greater review effort, including the need to clarify whether an AI agent can carry out its proposed actions. 

\textbf{When AI agents initiate code reviews, conversations end in 30.6\% fewer exchange rounds than when human reviewers review human-written code.} As shown in Figure~\ref{fig:rq2_all}, when AI agents initiate reviews, 85.2--86.7\% of conversations terminate at the AI agent's first review state without follow-up from the AI agent or human contributors. In contrast, in human review human-written code conversation, 48.9\% of cases involve multiple exchange rounds before ending. These results indicate that AI-agent-initiated reviews are typically less frequently developed into multi-turn discussions. 

\textbf{Human reviews of human-written code show a higher rate of related code changes without further human--AI exchanges than human reviews of AI-generated code, with 47.6\% for HRH compared with 17.5\% for HRA.} As shown in Fig.~\ref{fig:rq2_all}, the corresponding rates for AI-agent reviews are also lower than the HRH baseline, at 24.5\% for ARH and 30.9\% for ARA. Among ARH conversations reaching each successive AI-agent state, the proportions terminating in related code changes are 24.5\%, 32.8\%, and 35.3\%, respectively. However, the proportion of conversations ending with related code changes falls from 19.1\% after a human responds to the AI agent (i.e., HC $\rightarrow$ A $\rightarrow$ H $\rightarrow$ CC) to 7.2\% after another exchange between the AI agent and a human (i.e., HC $\rightarrow$ A $\rightarrow$ H $\rightarrow$ A $\rightarrow$ H $\rightarrow$ CC). In HRA conversations, the rate of related code changes decreases at later human states, while continued discussion after human responses becomes more frequent (66.4\% to 77.5\%). These results indicate that human reviews of agent-generated code involve more back-and-forth discussion between humans and AI agents than HRH when conversations are associated with related code changes.


\textbf{Cross-agent analysis shows that when Claude Code initiates a review, conversations involve significantly less back-and-forth discussion and have a lower rate of related code changes than reviews initiated by AI agents overall, with small effect sizes.} As shown in Table~A3 in the Appendix~\cite{online_appendix}, when Claude Code initiates a review, 7.3\% of conversations involve further discussion (\textit{>1C}), compared with 14.6\% overall. In addition, 22.5\% are followed by code changes, compared with 31.2\% overall. However, when Gemini Code Assist initiates a review, 19.8\% of conversations involve further discussion, significantly higher than 14.6\% overall (small effect size). The remaining differences are not statistically significant. These indicate that the overall results are generally consistent across AI agent reviewers, with small differences for reviews initiated by Claude Code and Gemini Code Assist.


\smallskip
\begin{Summary}{Summary of RQ2}{
When human reviewers review AI-generated code, conversations involve 11.8\% more exchange rounds than in reviews of human-written code. Moreover, human reviews of human-written code show a significantly higher initial rate of related code changes than reviews involving AI agents. 
}
\end{Summary}
\subsection{RQ3: \rqthree}\label{sec:RQ3}

\motivation In the second research question, we analyze human-AI interaction patterns and whether review conversations are followed by related code changes. However, this analysis does not show how frequently reviewers' explicit code suggestions are adopted or how code quality changes following their adoption. In this research question, we examine how frequently suggestions from AI agents and human reviewers are adopted and how code quality differs before and after adopted suggestions are applied.

\approach To compare suggestions from AI-agent and human reviewers, we measure how often their suggestions are adopted into the codebase. We then compare code metrics before and after each adopted suggestion is applied and categorize the reasons why AI-agent suggestions are not adopted.


\textbf{Measuring suggestion adoption.}\label{sec:code_suggestions} To understand how often review suggestions from human reviewers and AI agents are adopted into the codebase, we extract and analyze code suggestions from pull requests. Reviewers propose code suggestions as explicit code modifications within their comments, which developers can directly commit to the codebase~\cite{github_suggested_changes}.

Following the previous approach~\cite{bouraffa2025developersusecodesuggestions}, we statically extract code suggestions from pull requests by identifying code blocks enclosed in triple backticks followed by the word ``suggestion'' (\texttt{```suggestion}), and we identify code suggestions in each conversation within PRs. From 278,790 inline code review conversations, 113,684 conversations contain code suggestions. AI agents provide 88,011 code suggestions from a total of 155,397 conversations (56.6\%), while human reviewers provide 25,673~(20.8\%) code suggestions.

To determine whether each suggestion is merged into the production code, we follow an approach provided by Goldman et al.~\cite{goldman2025typescodereviewcomments}. We identify a code review comment as adopted if a subsequent commit within the same pull request modified the exact line where the code suggestion is placed. Line modification alone does not confirm adoption, as developers may change the same line for unrelated reasons. To filter false positives, we follow Svajlenko and Roy~\cite{codeclone}, who represent each code suggestion as a set of terms and measure similarity using the modified Jaccard similarity metric, where the similarity between two code fragments $f_1$ and $f_2$ is defined as $s(f_1, f_2) = \frac{|f_1 \cap f_2|}{\max(|f_1|, |f_2|)}$. A suggestion is labeled as adopted only if $s(\text{committed}, \text{suggested}) > s(\text{committed}, \text{original})$, indicating the committed code moved closer to the suggested code than to the original code. In addition, we verify that the modification is not reverted in subsequent commits and that the pull request is successfully merged, as unmerged or reverted changes do not persist in the codebase.

To compare how frequently human and AI agent suggestions are adopted, both overall and across feedback types, we define the adoption rate as the ratio of adopted suggestions to total suggestions for each reviewer type:
\begin{equation}
\text{Adoption Rate} = \frac{\text{\# Adopted Suggestions}}{\text{\# Total Suggestions}}
\end{equation}
We group code suggestions by feedback type using the taxonomy described in Section~\ref{sec:RQ1} and calculate the adoption rate separately for human reviewers and AI agents within each feedback type.

\textbf{Categorizing unadopted suggestions.} To understand why AI agent suggestions are not adopted, we examine review conversations where code authors or other reviewers responded to unadopted code suggestions from AI agents. We categorize the reasons for non-adoption by randomly sampling 383 unadopted code suggestions from AI agents with a 95\% confidence level and 5\% margin of error~\cite{ahmad2017determining}. We then use an LLM to provide an independent automated assessment, and the first author to review and finalize the labels through open coding~\cite{ahmed2025can,yu2026reusable}. The initial labels generated by GPT-4.1-mini achieve a Cohen's $\kappa$~\cite{cohen1960coefficient} of 0.76 when compared to the final human-verified labels, indicating substantial agreement. 

\begin{table}[t]
\centering
\caption{Suggestion adoption rates by feedback type. A ``-'' indicates insufficient data for that reviewer type. Adopt = suggestion adoption ratio. vs.~Human = difference in adoption rate (Agent Adopt - Human Adopt).}
\label{tab:rq3_feedtype_acc}
\begin{adjustbox}{width=\columnwidth}
\begin{tabular}{@{}l*{5}{r}@{}}
\toprule
\multirow{2}{*}{\textbf{Feedback Type}} & \multicolumn{2}{c}{\textbf{Human}} & \multicolumn{3}{c}{\textbf{Agent}} \\
\cmidrule(lr){2-3} \cmidrule(lr){4-6}
 & \textbf{\#Sugg.} & \textbf{Adopt} & \textbf{\#Sugg.} & \textbf{Adopt} & \textbf{vs.~Human} \\
\midrule
\textbf{\mbox{Knowledge Transfer}} & 2,038 & 75.8\% & - & - & - \\
\hline
\textbf{\mbox{Code Improvement}} & 20,166 & 59.2\% & 53,501 & 16.8\% & -42.3\% \\
\hline
\textbf{\mbox{Defects}} & 616 & 53.7\% & 33,019 & 16.7\% & -37.0\% \\
\hline
\textbf{\mbox{Understanding}} & 1,354 & 51.0\% & - & - & - \\
\hline
\textbf{\mbox{Misc}} & 1,499 & 1.2\% & - & - & - \\
\hline
\textbf{\mbox{External Impact}} & - & - & 658 & 6.5\% & - \\
\hline
\textbf{\mbox{Testing}} & - & - & 833 & 7.6\% & - \\
\midrule
\textbf{\mbox{Total}} & \textbf{25,673} & \textbf{56.5\%} & \textbf{88,011} & \textbf{16.6\%} & \textbf{-39.9\%} \\
\bottomrule
\end{tabular}
\end{adjustbox}
\vspace{-10pt}
\end{table}

\textbf{Assessing code metrics.} To examine whether suggestion adoption is associated with changes in code quality in terms of size, complexity, coupling, and cohesion, we compare the corresponding metric values before and after applying each adopted suggestion. We apply SciTools Understand~\cite{scitools_understand_metrics} to measure 111 code metrics before and after applying each adopted suggestion. These metrics capture properties of the source file, and the classes as well as functions defined within the source file, such as code size, complexity, coupling, and cohesion. 

To ensure that SciTools Understand can compute meaningful metrics, we restrict the analysis to suggestions with related code changes that modify source code. We exclude the adopted changes related to documentation and configuration files. Since multiple suggestions may have related code changes in the same source file, we assess each suggestion separately by applying only its related code changes to the baseline file while keeping all other content unchanged. Specifically, for each adopted suggestion, we retrieve the source file before the suggestion was applied and use it as the before-suggestion-adoption file. We apply only the diff hunk associated with the related code changes for the suggestion to create the after-suggestion-adoption file. All other code remains unchanged from the before-suggestion-adoption file. Our final analysis includes 3,467 adopted suggestions from human reviewers and 5,518 from AI-agent reviewers. 

To quantify the code-metric changes associated with each adopted suggestion, we follow Bouraffa et al.~\cite{bouraffa2025developersusecodesuggestions} to calculate the metric delta ($\Delta M$):
\begin{equation}
\vspace{-5pt} 
\Delta M = M_{\text{after}} - M_{\text{before}}
\end{equation}
\textit{where $M_{\text{before}}$ and $M_{\text{after}}$ represent the metric values before and after applying the code suggestion.}

Finally, we compare the metric deltas between suggestions from AI agents and human reviewers. Since the distributions are non-normal and we test 111 metrics simultaneously, we use the Mann-Whitney U test with Bonferroni correction~\cite{Hahn01101984}.



\textbf{The code suggestions proposed by human reviewers are adopted at a rate 39.9 percentage points higher than suggestions generated by AI agents.} As shown in Table~\ref{tab:rq3_feedtype_acc}, the highest adoption gap appears in the feedback type that shares project-specific knowledge (i.e., knowledge transfer), such as renamed classes and repository conventions. 
In contrast, AI agent reviewers rarely provide knowledge transfer suggestions (see Section~\ref{sec:RQ1}). Therefore, AI agent builders can consider designing AI agents to retrieve project-level knowledge related to the reviewed code, such as recently renamed classes and repository conventions, before generating suggestions.



\begin{table}[t]
\centering
\small
\caption{Reasons for unadopted AI code suggestions in a
statistically representative random sample ($N=383$)}
\label{tab:rq3_unadopted_categories}

\begin{tabularx}{\columnwidth}{
    @{}
    l
    @{\hspace{\tabcolsep}}
    >{\raggedright\arraybackslash}X
    @{\hspace{\tabcolsep}}
    r
    @{}
}
\toprule
\textbf{Category} &
\textbf{Definition} &
\textbf{Freq.} \\
\midrule

Incorrect &
The AI suggestion is incorrect or would break the code. &
28.7\% \\
\cmidrule(lr){1-3}

Alternative Fix &
The participant applies a different fix. &
24.0\% \\
\cmidrule(lr){1-3}

Not Needed &
The suggestion is unnecessary or already handled. &
16.4\% \\
\cmidrule(lr){1-3}

Claimed Fixed &
The participant claims the issue is fixed, but no corresponding code commit is found. &
11.2\% \\
\cmidrule(lr){1-3}

Developer Prefer &
The participant prefers a different implementation or coding style. &
10.2\% \\
\cmidrule(lr){1-3}

Deferred &
The participant plans to address the suggestion in future work. &
8.4\% \\
\cmidrule(lr){1-3}

Others &
The response is unclear or ambiguous. &
1.0\% \\

\bottomrule
\end{tabularx}
\vspace{-5pt}
\end{table}

\textbf{Among the sampled unadopted code suggestions from AI agents, incorrect suggested code is the most common reason for non-adoption, accounting for 28.7\% of cases.} As shown in Table~\ref{tab:rq3_unadopted_categories}, these suggestions would break the build or conflict with the project’s implementation. The second most common reason is that developers implement an alternative fix, accounting for 24.0\% of cases. Other reasons include AI agents proposing issues that have already been handled, conflicting with design decisions, or suggesting changes deferred to future work. 

\textbf{Adopted suggestions from AI agents are associated with significantly larger increases in code complexity and code size than adopted suggestions from human reviewers.} Table~\ref{tab:rq3_codemetrics_reviewer} compares the mean metric delta of adopted suggestions between human reviewers and AI agents. Suggestions from human reviewers are associated with significantly lower changes in six complexity metrics than suggestions from AI agents. The average differences between the before-suggestion-adoption and after-suggestion-adoption metric values of these six complexity metrics range from -0.074 to -0.006 for human reviewers and from 0.007 to 0.059 for AI-agent reviewers. Suggestions from AI agents also show larger increases in code size, with 0.241 compared with 0.087 in Code Lines, 0.093 compared with 0.033 in Executable Statements, and 0.010 compared with 0.003 in Functions. The results suggest that practitioners may need to carefully review adopted AI agent suggestions to avoid degrading code maintainability.


\begin{table}[t]
\centering
\caption{Code metric changes for adopted suggestions from human and AI-agent reviewers. $\Delta M$ denotes the before-to-after change. Only metrics differing significantly between reviewer types are shown.}
\label{tab:rq3_codemetrics_reviewer}
\resizebox{\columnwidth}{!}{%
\begin{tabular}{@{}llrr@{}}
\toprule
\textbf{Category} & \textbf{Metric} & \textbf{Human} & \textbf{Agent} \\
 & & \textbf{Mean $\Delta$} & \textbf{Mean $\Delta$} \\
\midrule
\textbf{Complexity} & Sum Strict Modified CC & $-$0.074 & 0.059 \\
 & Sum Modified CC & $-$0.056 & 0.045 \\
 & Sum CC & $-$0.063 & 0.044 \\
 & Average Strict Modified CC & $-$0.011 & 0.010 \\
 & Average Modified CC & $-$0.007 & 0.007 \\
 & Average CC & $-$0.006 & 0.007 \\
\midrule
\textbf{Code Size} & Executable Statements & 0.033 & 0.093 \\
 & Code Lines & 0.087 & 0.241 \\
 & Functions & 0.003 & 0.010 \\
\bottomrule
\end{tabular}%
}
\vspace{1pt}
{\scriptsize\raggedright CC = Cyclomatic Complexity.\par}
\vspace{-10pt}
\end{table}



\textbf{Adopted suggestions generated by Greptile show a decrease in cyclomatic complexity that differs significantly from the increase observed for adopted AI agent suggestions overall, with a mean change of -0.979 versus 0.045.} As shown in Table~A4 in the Appendix~\cite{online_appendix}, Greptile-generated adopted suggestions also show decreases in the number of declarative statements and average lines of code that differ significantly from the increases observed for adopted AI agent suggestions overall. Moreover, adopted suggestions generated by Cursor and Sourcery AI show decreases in the number of executable statements that differ significantly from the increase observed for adopted AI agent suggestions overall. The remaining agents (i.e., Copilot, CodeRabbit, Gemini Code Assist, and Claude Code) show no significant differences from the overall AI-agent distribution across the 111 code metrics.

\smallskip
\begin{Summary}{Summary of RQ3}{Suggestions proposed by human reviewers have a significantly higher overall adoption rate than suggestions proposed by AI agents. When adopted, suggestions from AI agents show significantly larger increases in complexity and code size metrics.}
\end{Summary}
\section{Implication}\label{sec:implication}
In this section, we discuss the implications of our study for the practitioners and tool makers.

\textbf{Bridging the socio-technical gap between AI agents and human reviewers in code review.} Our results show that human-reviewer feedback focuses more frequently on understanding and clarifying developer intent, while these feedback types appear less frequently in AI-agent reviews. Conversations in which humans review human-written code show a higher rate of related code changes than conversations involving AI agents. These findings suggest that mechanisms for requesting clarification and retrieving project-specific context may help AI agents address intent-related feedback. We therefore encourage researchers to develop methods that enable AI agents to infer and articulate developer intent, supporting effective human-agent collaboration.

\textbf{Investigating project-specific context in AI-agent suggestions.} AI-agent reviewers generate more code suggestions than human reviewers but have lower adoption rates. Among the sampled unadopted AI-agent suggestions, 28.7\% are incorrect and 24.0\% receive developer-chosen alternative fixes. These non-adoption patterns may reflect differences in suggestion relevance, presentation format, access to project-specific information, or developer preferences. Future studies could evaluate whether providing build configurations, prior review decisions, and other project-specific information changes the suggestion accuracy and adoption rates.



\textbf{Improving readability and clarity of AI-agent feedback.} Our results show that human-reviewer comments have significantly higher reading-ease scores than AI-agent comments. AI-agent comments also more frequently combine an evaluation of a code weakness with a proposed fix in the same comment. Tool builders can consider evaluating AI-agent feedback not only for technical coverage but also for readability and clarity.

\textbf{Supporting agent responsiveness across multi-turn review interactions.} Our results show that conversations in which humans review AI-generated code contain the most comments on average. However, conversations in which humans review human-written code show a higher rate of related code changes than conversations involving AI agents. This pattern may reflect challenges in how AI agents maintain context and incorporate reviewer feedback across multiple exchanges. We encourage tool builders to investigate whether stronger context tracking and self-verification can support AI-agent responses to reviewer concerns across multiple exchanges.


\section{Threats to Validity}
In this section, we examine the potential threats to the validity of our study.

\textbf{Threats to Construct Validity.} 
Our study focuses on AI agents whose identities are explicitly disclosed through their account names. Our comparison of human and AI-agent behavior in the review process is based on account identity. However, pull requests and review comments posted by human accounts may contain content generated or modified using AI tools. These contributions remain in the human category, including cases where authors disclose AI assistance. GitHub metadata does not capture such assistance, which limits our ability to distinguish human-written from AI-assisted content within this category. Robbes et al.~\cite{agent_much} similarly acknowledge that AI agent activity can remain unobserved when attribution traces are disabled or developers manually commit agent contributions. This limitation may affect the estimated differences between the human and AI-agent categories.


\textbf{Threats to Internal Validity.} The assessment of code quality uses metrics of code size, complexity, coupling, and cohesion at the source-file level. These metrics do not capture project-level architectural changes, security vulnerabilities or fault proneness. In addition, the differences in code metric changes associated with adopted suggestions from human and AI-agent reviewers may be influenced by repository characteristics, programming language, pull request size, code churn, developer experience, or the AI review tool used.

\textbf{Threats to External Validity.}
AI agents developed by different vendors may use different underlying models, tools, and interaction designs, which may lead to product-specific review behavior. To assess this threat, we conduct a cross-agent analysis of seven agents from different vendors across all three RQs. Although most differences are not statistically significant or have negligible-to-small effect sizes, several product-specific differences have emerged. In particular, Cursor is the main outlier in review-comment characteristics, Claude Code and Gemini Code Assist show small differences in interaction patterns, and suggestions generated by Greptile, Cursor, and Sourcery AI show different changes in some specific code metrics. Moreover, agent version information is unavailable in GitHub review comments, so our analyses cannot account for differences across versions of the same agent. Therefore, other agents or product versions may yield different results.

\section{Related Work}\label{sec:related_work}

In this section, we review the literature related to code review by human reviewers and AI-based code review.

\textbf{Human code review.} Code review is a fundamental practice in software development where developers examine code changes before merging. Prior studies~\cite{6606617,review_Dynamics, googlecodereview2018} have extensively investigated reviewer expectations, feedback quality, and review outcomes.

Bosu et al.~\cite{review_Dynamics} survey code review practices across open-source and industrial settings, finding that reviewer experience and prior engagement with the codebase increase the likelihood of providing useful feedback. Bouraffa et al.~\cite{bouraffa2025developersusecodesuggestions} study how developers use code suggestions in pull request reviews, finding that the use of code suggestions positively affects the merge rate of pull requests. Goldman et al.\ \cite{goldman2025typescodereviewcomments} analyze which types of human-written review comments developers most frequently resolve. 


\textbf{AI-Based Code Review.} With the emergence of generative AI, recent studies have investigated the roles of AI agents as both pull request authors and code reviewers. For AI-generated PR descriptions, Xiao et al.\ \cite{Xiao_2024} examine their adoption and related developer interventions, while Gong et al.\ \cite{Gong2026messagecode} assess the consistency between PR descriptions and their underlying code changes. Gonzalez and El Mezouar \cite{Gonzalez2026BotRej} and Abujadallah et al.\ \cite{Abujadallah2026RejectioPR} investigate why agent-authored PRs are rejected and quantify the associated code churn and review effort. Nachuma and Zibran \cite{Nachuma2026Codereview} examine the integration of agent-authored PRs and find that reviewer engagement is associated with successful integration. For AI agents acting as code reviewers, Sun et al.\ \cite{sun2025doesaicodereview} and Olewicki et al.\ \cite{olewicki2024impactllmbasedreviewcomment} evaluate the adoption, effectiveness, and acceptance of AI-generated review comments. Zhong et al.\ \cite{zhong2026humancentricagenticcodereview} provide a longitudinal perspective by examining how AI reviewer adoption patterns evolve at both the project and pull request levels.



In this study, we shift the focus from isolated AI-generated artifacts to the multi-turn collaborative dialogue between humans and agents. We characterize the behavioral differences across four reviewer pairings (HRH, HRA, ARH, ARA) and model the negotiation process using interaction sequences to capture multi-turn dynamics. Finally, we connect these interaction patterns to review outcomes by examining suggestion adoption and changes in code metrics associated with adopted suggestions from human and AI-agent reviewers.
\section{Conclusion}\label{sec:conclusion}

In this study, we analyze 278,790 code review conversations across 300 open-source GitHub projects to understand how AI agents and human reviewers differ in feedback characteristics, interaction patterns, and suggestion quality. We find that AI-agent comments focus primarily on Code Improvement and Defect Detection and have lower reading-ease scores, while human reviewers provide broader feedback. Conversations in which humans review human-written code show a higher rate of related code changes than conversations involving AI agents. Human-reviewer suggestions have higher adoption rates, while adopted AI-agent suggestions show larger mean increases in selected complexity and code-size metrics. For future work, we aim to improve AI agent context awareness, explore multi-agent mechanisms for sharing project knowledge, and reduce incorrect suggestions to bridge the quality gap between human and AI agent code review.
 


%
%

\balance
\bibliographystyle{IEEEtran}
\bibliography{Bib}   

\end{document}